\documentstyle[epsf,prl,twocolumn,aps]{revtex}
\begin{document}
%
%
%
\title{
Spin- Flip Torsion Balance}
\author{Peter Fulde and Stefan Kettemann}
\address{ Max- Planck Institut f\" ur Physik Komplexer Systeme\\
 N\" othnitzerstr. 38, 01187 Dresden, Germany}
\date{\today}
\maketitle

 \begin{abstract}
 The spin flip of the conduction electrons at the interface of a ferromagnetic
 and a nonmagnetic part of a metallic wire,
  suspended between two electrodes, 
 is  shown to tort the wire when a current is driven through it.
 In order to enhance the effect it is suggested to 
 use an alternating current in resonance with the torsional 
 oscillations. 
  Thereby the  magnetic polarization of the conduction
 electrons in the ferromagnet can  be measured directly,
 and compared to the total magnetization.
  This may yield new information on the 
 transport properties  of the narrow band
 electrons in itinerant ferromagnets.   
 
Pacs- numbers:  75.10.Lp,75.50.Bb,75.80.+q
\end{abstract}
%
%
%
\maketitle
%

 \begin{figure*}[bhp]\label{fig1}
 \centerline{\hbox{\epsfbox{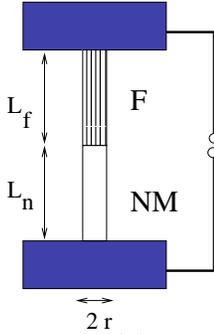}}}
 \caption{ A ferromagnetic (F) and a nonmagnetic (NM)
 metallic wire are suspended between two  electrodes}
 \end{figure*}

 Torsion angle measurements 
 have achieved  a high sensitivity  
 and   resulted in 
 the early measurement of the gyromagnetic ratio of electrons
 \cite{rich,einstein}, the
 experimental  proof 
 of the quantization of the angular momentum of 
 circularly polarized photons\cite{beth}, and the experimental proof of the
 flux quantization through a superconducting loop\cite{doll}. 
 
  Here we propose to measure the torsion angle 
 resulting from the spin flip of conduction electrons at the
 interface of a ferromagnetic and a paramagnetic metallic wire.
  This may shed new light on the transport 
 properties of the d- electrons 
  in ferromagnetic transition metals like Fe, Co or Ni\cite{aha}.

\section{Spin Flip Induced Torsion}
 Let us consider a wire whose one end is made of a ferromagnet, the
 rest being a nonmagnetic metal, see Fig. 1, 
 which is fixed rigidly to the electrodes.
 Ideally, the ferromagnet is uniformally
 spin  polarized in a direction parallell
 to the wire.
  If a current is driven through it, the 
 spin polarized electrons in the ferromagnet loose 
 on average
each an angular momentum
 $ (\hbar/2) \Pi$ at the interface between the two parts of the wire.
 Here, $\Pi$ is the ratio of conduction electrons 
 which flip their spin when they enter or leave the ferromagnet.
 We assume that the 
spin flip of the conduction electrons 
occurs in a region of length $l$
small compared to the length $L$ of the wire.
  Thus, a current $I$ gives rise to a rate of spin flip 
 $\Pi I/q$, and 
 may thus produce a maximal torque 
along the wire of magnitude
\begin{equation}\label{flip}
 D_z = \frac{1}{2} \hbar \hspace{.1cm} \Pi \hspace{.1cm} \frac{I}{q},
\end{equation}
 where $q$ is the  charge of an electron.

\section{Torsion Balance}
 The torque caused by the finite spin flip rate, Eq. ( \ref{flip} ),
is in balance with the torque due to the 
  elastic torsion  of the wire, as well as a
 torque due to the finite inertia 
 of the wire, and a  friction torque when the torsion 
angle changes in time.
 
 A torsion of the ferromagnetic wire by an angle
$\alpha$ causes 
 a torque  of magnitude \cite{einstein,landau} 
\begin{equation}
D_{elastic} = \frac{\pi}{2} G_f \frac{r^4}{L_f}  
\alpha,
\end{equation}
where $G_f $ is the shear modulus, $ L_f$ the length
 of the ferromagnetic wire and $r$ its radius.
 We assume that the torque on the nonmagnetic wire 
 when it is torted by the angle $\alpha$,
 is smaller than
 the torque
in the ferromagnetic wire $D_{elastic}$, so that the torsion angle $\alpha$ 
is determined by $D_{elastic}$, only.  
 Note that we consider here the situation when the 
 wires are under no tension.
  If there is a force $F$ on the wires,
 producing the tension $F/(\pi r^2)$, it 
 has to be substituted for 
 the shear modulus $G_f$.

 When the angle $\alpha$ changes in time,
  there
  is an additional  torque due to the inertia $J$ of the wire, 
\begin{equation}
D_{inertia} = J \hspace{.1cm} \ddot \alpha,
\end{equation}
  The inertia of the two wires, each fixed at one end, 
is 
\begin{equation}\label{inertia}
J = J_0 + \pi \rho_f r^4 L_f/4 + \pi \rho_n r^4 L_n/4,
\end{equation}
 where the subscript $n$ 
 denotes the values for the nonmagnetic wire,
 and $\rho_f, \rho_n$ is the mass  density of the respective wire.
 Note that the inertia may be changed 
independently of the wire geometry 
  by attaching a weight. Therefore  we have included another finite 
contribution $J_0$ to the  inertia. 

 There is also a small friction torque, which can be 
 taken of Stokes form:
\begin{equation}
D_{friction} = R \hspace{.1cm} \dot \alpha,
\end{equation}
 where $R$ is the friction constant.

 Thus, in  total there is a balance of torques,
\begin{equation}\label{balance}
D_z = D_{elastic} + D_{inertia} + D_{friction},
\end{equation}
 yielding a second order differential equation 
 for the torsion angle $\alpha$.

 If the current $I$ is constant in time, 
   the  torsion will reach equilibrium at a finite  angle 
\begin{equation}\label{const}
\alpha = \frac{1}{\pi} \frac{L_f}{G_f  r^4} 
\frac{I}{q}  \hbar \hspace{.1cm} \Pi.
\end{equation}
 
 In order to enhance the torsion angle 
  one can send  an alternating current through 
 the wire and bring it in resonance with the 
 eigen frequency of torsion  oscillations of the wire system.
  For a sinusoidal  current, $ I = I_0 \sin ( \omega t ) $,
 there is resonance at the  frequency,
\begin{equation} \label{freq}
 \omega = \sqrt{ \frac{G_f}{\rho_f} } g \frac{1}{L_f}. 
\end{equation}
  
 In resonance the maximal torsion angle 
 is limited by the  friction, and there is a torsional oscillation
 $\alpha(t) = \alpha_0 \cos ( \omega t )$ 
 where the amplitude is obtained
 from Eq. (\ref{balance}) as, 
\begin{equation}
\alpha_0 = - \frac{1}{2 R \omega} \frac{I_0}{q}  \hbar
\hspace{.1cm}  \Pi. 
\end{equation}

 The friction constant $R$ is related to the decay rate $\Gamma$  
 with which a torsion oscillation  decays, when there is no
 driving force, $I=0$:
$\alpha(t) = \alpha_0 \cos ( \omega t ) \exp ( - \Gamma t)$.

 Substitution in  Eq. ( \ref{balance} ) gives 
 $\Gamma = R/(2 J)$. In order that resonance can be observed,
  the decay rate must be much smaller than the resonance frequency,
 $\Gamma << \omega$.

 Using Eq. ( \ref{freq} ) for the resonance frequency
 and Eq. ( \ref{inertia} ) for the inertia,
 we obtain  the maximal torsion angle as
\begin{equation} \label{a0}
\alpha_0 = - \frac{1}{2 \pi}   
\frac{1}{ \Gamma} \frac{1}{ r^4} \frac{1}{\sqrt{ \rho_f G_f}}
 g \frac{I_0}{q}  \hbar
\hspace{.1cm}  \Pi,
\end{equation}
where
\begin{equation}
g^2 = 2/( 1 + \frac{\rho_n}{\rho_f}  \frac{L_n}{L_f} + 4 
\frac{J_0}{\rho_f r^4 L_f}).
\end{equation}
  Note, that the factor g depends on the dimensions of the wires 
and the additional inertia $J_0$,
and is on   the order of $1$.
  
 Thus, when $\Gamma$ is obtained  from a relaxation  
 measurement, the maximum amplitude of the torsion angle in resonance,
 Eq. (\ref{a0}), is a  direct measure of the polarization of the 
 conduction electrons $\Pi$. 
  
\section{Thermal Fluctuations}
 The torsion due to the spin flip rate 
 has to be compared to thermal fluctuations 
 of the torsion angle, $\delta \alpha$, at the temperature $T$ of the wire.
 Using Boltzmann's equipartition theorem,
 setting the elastic energy $E(\delta \alpha)$ equal to 
$ k_B T/2$,
 where \cite{landau}
\begin{equation}
E(\delta \alpha ) = \int_0^{\delta \alpha} d \alpha D_{elastic} ( \alpha ) =  
 \frac{\pi}{4}  (\delta \alpha)^2  G_f \frac{r^4}{L_f},
\end{equation}
 we find that the torsion angle may maximally fluctuate by 
\begin{equation}
\delta \alpha = \sqrt{\frac{2}{\pi}} \frac{1}{r^2} \sqrt{ \frac{k_B T L_f}{G_f} }.  
\end{equation}

\section{Example}

 As an example we consider at room temperature $ T = 273 K$
 a ferromagnetic  wire made of  $Fe$
 with a Curie temperature of $T_c = 1043 K$, 
 a torsion modulus  $G_f = 84 * 10^9 kg/(m s^2)$,
 and density  $\rho_f = 7.86 * 10^3 kg/m^3$\cite{ashcroft}.    
  The geometric parameter is chosen as $g =1$.
 In the following we
 write the physical parameters in experimentally relevant units:
 the  current $ I = {\bf I} \hspace{.2cm} A = {\bf I}
\hspace{.2cm}  6.25 * 10^{18} q/s $,
 the radius, $r = {\bf r} \hspace{.2cm} 10^{-6} m $,
 the length $L_f = {\bf L_f} \hspace{.2cm}  m$,
 and the decay rate 
$\Gamma = {\mbox {\boldmath$ \Gamma$}} \hspace{.2cm} Hz$
 where the 
 letters in bold face
 denote the dimensionless parameters.
 Thus,  we obtain for the torsion angle Eq. ( \ref{const} )
 as a function of only dimensionless parameters:
\begin{equation}
\alpha = 2.5 \frac{{\bf L_f} }{{\bf r}^4}
 \hspace{.1cm} {\bf I} \hspace{.1cm} \Pi  * 10^{- 3}. 
\end{equation}
In resonance the maximum torsion angle is 
 from Eq. ( \ref{a0} ) 
\begin{equation}
\alpha_0 = 4.10 \frac{1}{{\bf r}^4} \frac{1}{
  {\mbox {\boldmath$ \Gamma$}}} 
{\bf I_0} \Pi 
\end{equation}
 with the resonance frequency
\begin{equation}
\omega = 3.27 * 10^3  \frac{1}{\bf L_f} Hz.
\end{equation}
 As an example, when a constant current of
 $I = 1  A$ is driven through an iron wire of 
 thickness $20 \mu m$ and length $1 m$, 
 it becomes torted by  the angle  $2.5 \hspace{.2cm}  \Pi * 10^{-7}$
 while a
 maximum torsion angle
$\alpha_0 = 4.10  \hspace{.2cm} \Pi * 10^{-4}$ 
 can be achieved in resonance 
 with an oscillating current of amplitude $I_0 = 1  A$, 
 with a frequency of $3.27 *  10^3 Hz$, if the system has an
 independently measured  relaxation rate
 $\Gamma = 1 Hz$. 

 For the thermal torsion angle fluctuations
 we obtain at a temperature $T = {\bf T} K $:
\begin{equation}\label{fluc}
 \delta \alpha = \frac{\sqrt{ {\bf T} {\bf L_f} }}{{\bf r}^2} 1.02 * 10^{- 5}
\end{equation}
 which gives at room temperature $ T =  273 K$ for the above example 
a  fluctuation
of  $\delta \alpha = 1.69 * 10^{-6}$.

 To estimate the heating of the wire with respect to the 
environment with constant temperature $T$, we assume that the heat current is 
 radially emitted out of the wire. 
 With the 
values for $Fe$ for the
 resistivity $\rho_e = 8.9 * 10^{-6} \Omega cm$
 and the thermal conductivity $\kappa = 80 J/( s K m)$,
 we obtain that the radial temperature gradient 
 is 
\begin{equation}
\Delta T = {\bf I}^2 \frac{1}{{\bf r}^2} 56  K,
\end{equation} 
  For the example follows a heating of $.56  K$,
 which can be neglected at room  temperature.
 
\section{Conclusions}
 
   The spin flip rate at the interface between a ferromagnetic and a
 nonmagnetic wire results in a torsion of the wire. 
  The magnitude of the torsion angle seems, at least in resonance, in 
 an experimentally accessible range, and can even
 at room temperature exceed the   thermal
 torsion  fluctuations. The effect can be strongly enhanced 
 and its ratio
 to thermal fluctuations be increased, by decreasing the wire thickness,
 while its dependence on temperature is relatively weak.

 The torsion angle can therefore serve as a direct measure
 of the degree of magnetic polarization of the conduction electrons $\Pi$.
  $\Pi$ is known to be close to $1$ in doped $La Mn O_3$,
since the spin- up and spin-  down bands are well separated in energy
 in these compounds\cite{mn}. 
 In ferromagnetic transition metals
 like $Fe, Co, Ni$, 
 it is not yet known 
to which  degree  the $d- electrons$ contribute to the electric transport
 \cite{fulde}, and therefore the measurement
 of the degree of magnetic polarization of the conduction electrons
 $\Pi$ may lead to a better understanding of 
 that point.

 S. K. thanks Mukul Laad and
Mark Leadbeater for usefull discussions.

\end{document}